\documentclass[a4paper,twocolumn]{article}
\usepackage[latin1]{inputenc}
\usepackage{amsfonts}
\usepackage{latexsym}
\usepackage{amssymb}
\usepackage{graphicx} 
\title{Energy conserving Anisotropic Anhysteretic Magnetic Modelling for
  Finite Element Analysis}
\author{Jens~Krause\thanks{The author is with Rheinamic GmbH, Bornheimer
    Str. 33a, D-53111 Bonn, Germany; E-mail: jxkrause@arcor.de}}
\date{13 Dec 2012}
\begin{document}
\maketitle
\begin{abstract}
  To model ferromagnetic material in finite element analysis a correct
  description of the constitutive relationship (BH-law) must be found
  from measured data. This article proposes to use the energy density
  function as a centrepiece. Using this function, which turns out to
  be a convex function of the flux density, guarantees energy
  conservative modelling. The magnetic field strength can be seen as a
  derivative with respect to the flux density. Especially for
  anisotropic materials (from lamination and/or grain orientation)
  this method has advantages. Strictly speaking this method is only
  valid for anhysteretic and thermodynamically stable material.
\end{abstract}
\section{Introduction}
\label{sec:intro}
This essay discusses one aspect of numeric modelling of magnetic
fields: anisotropic magnetic material properties. In practical
electromagnetic applications anisotropic materials are common as
laminated electric steel with and without grain orientation. In the
design process finite element analysis (FEA) is widely used and
correct modelling of the material properties is essential. The
literature is wide spread, but considering saturation effects 
this essay contributes modelling that does not conflict with energy
conservation.

Application with use of electric steels are transformers, motors, and
generators. The main reason for lamination is that eddy currents in
stacking direction are suppressed. This lamination also has the
effect that the magnetic properties are anisotropic. Two different
types of electric steel are used: grain-oriented and
non-grain-oriented steels. The non-grain-oriented sheets have magnetic
properties that are isotropic in the sheet plane, whereas the
non-grain-oriented steels have different magnetic properties in
rolling direction and in transverse direction.

The discussion here is only applicable for perfect soft material,
i.e. material which has a nonlinear BH-characteristic but has no
hysteresis (meaning that the area of the hysteresis loop is
empty). This is the requirement that there exists a one-to-one
relationship between the vectorial flux density~$\bf B$ and the
vectorial magnetic field~$\bf H$. Models for hysteresis
(Jiles-Atherton, Preisach, and other) are not considered here.

The centrepiece of this essay is the use of the magnetic energy as a
function of the flux density as the centrepiece of the description of
magnetic properties. The magnetic field strength is in this concept
the gradient (with respect to the flux density) of the energy
density. The energy density can be used to describe linear and
nonlinear as well as isotropic and anisotropic material. It will be
used to derive equations for lamination effects and as a base for the
interpolation rule that extends measured data of grain-oriented steel
to a full 3-D description.

Also the energy density function will be used to graph
BH-characteristics by plotting lines of equal energy (iso-lines).

\section{Maxwell-Equations}
In order to clarify the notation, the relevant equations for
magnetostatics are repeated. The flux density is described by a
3-vector~$\bf B$ which is divergence free (Gauß-law):
\begin{eqnarray*}
  \label{eq:gauss}
  \nabla \cdot {\bf  B} = 0.
\end{eqnarray*}

The material properties are in the form of a function that maps~$\bf B$ to
the magnetic field~$\bf H$ which is also a 3-vector. Formally:
\begin{eqnarray}
  {\bf H} = F \left ( {\bf B} \right ).
  \label{eq:aniso-law-general}
\end{eqnarray}
Ampere's -law relates the magnetic field to the electric current
density vector~$\bf j$, such that the curl of the magnetic field is
equal to the current density:
\begin{eqnarray*}
  \label{eq:biot-savart}
  \nabla \times {\bf H} = {\bf j}.
\end{eqnarray*}
Many of the commonly used FEA software uses the vector potential~$\bf
A$ as the unknown function, which is linked to the flux density by 
\begin{eqnarray*}
{\bf B} &=& \nabla \times {\bf A}\\
0 &=& \nabla \cdot {\bf A}.
\end{eqnarray*}
The topic in this text is to discuss anisotropic materials which are
described with a general nonlinear function $F$. 

The differential reluctivity~$\nu$ is a 3-by-3 matrix and defined as
the partial derivative of the field strength with respect to the flux
density. The components are
\begin{eqnarray*}
\nu_{ij}=\frac{\partial H_i}{\partial B_j}.
\end{eqnarray*}

\section{Literature review}

Other authors often express the constitutive relation by employing the
effective reluctivity matrix (or equivalently the effective
permeability) in the form ${\bf H}=\nu_{eff}{\bf B}$, where
$\nu_{eff}$ in general depends on the field quantities.

The case of lamination is often treated in the following way
(eg. \cite{Silva1995}): assuming
steel has the material law
\begin{displaymath}
  {\bf H}=\left(
\begin{array}{ccc}
\nu_{xx}&0&0\\
0&\nu_{yy}&0\\
0&0&\nu_{zz}\\
\end{array}
\right){\bf B}
\end{displaymath}
and the lamination fill factors are $f_1$ for insulation and $f_2$ for
steel. The BH-relationship for the composite is
\begin{displaymath} 
  {\bf H}=\left(
    \begin{array}{c}
      \frac{B_x}{f_1\mu_0+f_2/\nu_{xx}}\\
      \frac{B_y}{f_1\mu_0+f_2/\nu_{yy}}\\
      B_z(\frac{f_1}{\mu_0}+\nu_{zz})
    \end{array}
  \right){\bf B}.
\end{displaymath}
But this relationship is only correct if the reluctivity is
constant. In reality the reluctivity must depend on
the flux density and the reluctivity function of the underlying steel
must be evaluated in the model using the flux density in the magnetic
material (and not the macroscopic value). The reference \cite{Kaimori2007}
gives a correct model, where one additional equation must be
solved. This agrees with the solution presented here, but is only
valid for isotropic material.

When modelling the in-plane anisotropy of grain-oriented electric
steel often only a limited number of sets of measure data (usually in
rolling and transverse direction) is available. Then interpolation
models must be found for intermediate directions. Usually a 2-D model
is often presented in literature with the constitutive relation
\begin{displaymath}
  {\bf H}=\left(\begin{array}{cc} \nu_{xx} & \nu_{xy} \\
      \nu_{yx}&\nu_{yy}\end{array}\right){\bf B}.
\end{displaymath}
Such an effective reluctivity is not uniquely defined. There are two
physically measurable degrees of freedom ($H_x, H_y$), but the
symmetric reluctivity tensor contains three values
($\nu_{xx},\nu_{xy}=\nu_{yx},\nu_{yy}$). So when determining the
reluctivity tensor a gauge has to be chosen, but no author actually
discusses the implication of such a choice
(eg. \cite{Enokizono1997,Shirkoohi1994,DiNapoli1983,Liu1994}). Without
going into the detailed equations the models all represent
BH-relationships with non-symmetric differential reluctivity, and
therefore violate energy conservation as detailed below.

In cases when measurements are available in the entire $B_x,B_y$-plane the
interpolation can be done directly on this data (\cite{Janicke1997}).

The energy density function (or co-energy density function) is  only
used by few authors (eg.  \cite{Silvester1991,Pera1993}). The results
presented here can be seen as en extension of these ideas so that the
energy density forms the centrepiece of all treatment of
BH-relationships and with special attention to anistropy from
lamination and grain-orientation.

\section{General ansiotropic BH-functions}
\label{sec:gen}

In the introduction the BH-function (eq.~\ref{eq:aniso-law-general})
was considered to be a general vector function. Now 
the properties of this function are discussed. 

Let $\gamma$ be a path in the space of $\bf B$,~i.e. a smooth function
$[0,1]\rightarrow R^3$. The energy density $w$ related with this path
is defined as
\begin{eqnarray*}
  w(\gamma)=\int_{t=0}^1 {\bf H}\cdot\gamma^\prime dt.
\end{eqnarray*}
This energy is stored in the magnetic field at one material point.
Energy conservation now means that $w$ depends on the start and end
points alone and not on the path.  Using zero as the start point 
the energy density can be defined as a function of $\bf B$
without referring to a particular path:
\begin{eqnarray*}
  w({\bf B})=\int_{B'=0}^{\bf B} {\bf H}\cdot{\rm d}{\bf B}.
\end{eqnarray*}
The following statements follow from the existence of such an energy
density function.
\begin{itemize}
\item The magnetic field is the gradient of the energy density
  \begin{eqnarray*}
    {\bf H} = \nabla_{\bf B}w
  \end{eqnarray*}
\item Since the differential reluctivity tensor is the second
  derivative of $w$ and is symmetric
  \begin{eqnarray*}
\nu_{ij}=\frac{\partial H_i}{\partial B_j}=\frac{\partial^2w}{\partial
  B_i\partial B_j}
  \end{eqnarray*}
must be fulfilled.
\end{itemize}
\section{Convex energy function}
\begin{figure}
  \input{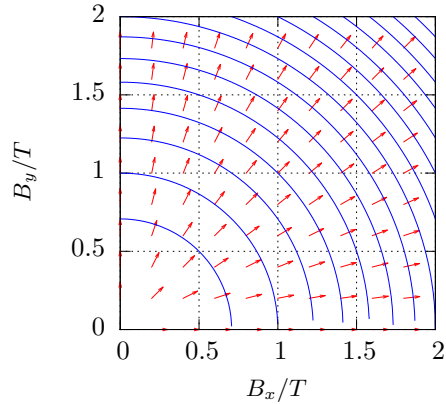}
  \caption{\label{fig:contour_iso}Contour lines of the energy density
    function for vacuum and direction of magnetic field strength}
\end{figure}

For isotropic BH-characteristics it obvious that $|\bf H|$ is
monotonously rising with $|\bf B|$. In consequence the derivative (the
differential reluctivity) is positive. Ref.~\cite{LLem} show that for
a thermodynamic stable material the differential permeabilty (and its
inverse the differential reluctivity) must be positive.  In an
anisotropic case the reluctivity is a symmetric matrix and the
requirement for positiveness is generalised as follows. A matrix $M$
is called {\em positive definite} if for all vectors $v$ the product
$v\cdot M v$ is always positive. It is therefore required that the
differential reluctivity is such a positive definite matrix.


The requirement for a positive definite second derivative can be linked
with the term of a {\em convex} function. A scalar function $f:
R^n\rightarrow R$ is called convex if for all $x,y\in R^n$ and all
$t\in [0,1]$ the inequality
\[
f((1-t)x+ty)>(1-t)f(x)+tf(y)
\]
holds. In words: the straight line connecting two points in the graph
of the function lies {\em above} the function. The word {\em convex}
can be explained, when looking at point sets like
\begin{eqnarray}
U(Q)=\{x\in R^n|f(x) \leq Q\}
\end{eqnarray}
for $Q\in f(R^n)$. For convex functions these sets are convex in the
ordinary sense. A set $U(Q)$ includes all points where the function
takes values smaller or equal than $Q$. An appropriate way of
displaying such functions is to look at iso-lines (where $f(x)=Q$),
because theses iso-lines enclose convex areas or volumes.

It can be demonstrated that for moderate additional assumption that
convex functions have positive definite second derivatives and vice
versa (see~\cite{BarnerFlohr}). 

The requirement for a positive definite reluctivity is also mentioned
in \cite{VandeSande2004}. It has the benefit that it guarantees a stable
finite element analysis.


\section{Examples}
\label{sec:examples}
\begin{figure}
  \input{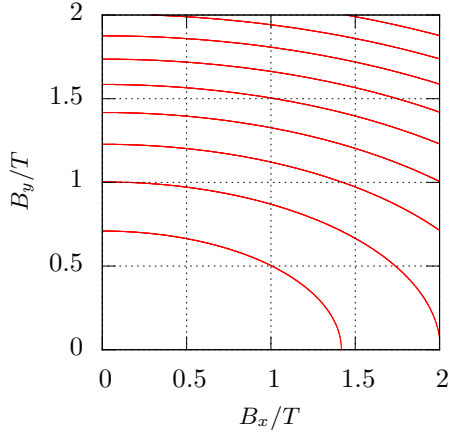}  
  \caption{\label{fig:contour_aniso}Contour lines of the energy
    density function for a linear anisotropic material. 
}
\end{figure}

For simple cases the energy density functions are as follows. A
natural way of graphing the energy function in 2-D planes is to use
contour lines of equal energy. These lines enclose convex areas if the
energy function is valid.

\begin{figure}
  \input{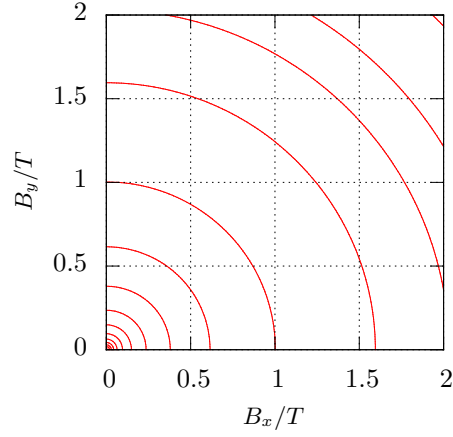}
  \caption{\label{fig:contour_steel_iso}Contour lines of the energy
    density function for mild steel.}
\end{figure}

\begin{itemize}
\item In vacuum:
  \begin{eqnarray}
\label{eq:energy_vacuum}
    w({\bf B}) =\frac{\bf B^2} {2\mu _{0}} 
  \end{eqnarray}
with $\mu_0=4\pi 10^{-7}\textrm{ Vs/Am}$.
As shown in fig.~\ref{fig:contour_iso} lines of constant energy
density are concentric circles around zero. This energy leads to the
magnetic field strength:
\begin{eqnarray}
  \label{eq:vacuum}
  {\bf H} =\frac{\bf B} {\mu _{0}} .
\end{eqnarray}
The mentioned figure also displays the direction of the field strength
$\bf H$ and shows that the field strength is perpendicular to isolines
of the energy density.
  \item Anisotropic linear case:
    \begin{eqnarray*}
      w({\bf B}) ={\bf B}\cdot\nu{\bf B},
    \end{eqnarray*}
where $\nu$ is the symmetric reluctivity matrix.
The contour lines are concentric ellipses
(fig.~\ref{fig:contour_aniso}) and the BH-law is
    \begin{eqnarray*}
      {\bf H} =\nu{\bf B}.
    \end{eqnarray*}
  \item Nonlinear isotropic case: the energy density only depends on
    the modulus of the flux density 
    \begin{eqnarray}
      \label{eq:energy_iso}
      w({\bf B}) =w_{iso}(|{\bf B}|).
    \end{eqnarray}
    The contour plot is similar to the plot for vacuum, only the
    spacing between the lines is different
    (fig.~\ref{fig:contour_steel_iso}). The underlying BH-relationship
    can be measured and is displayed in fig.~\ref{fig:bh-measurement}
    The resulting BH-relationship is
\begin{eqnarray}
  \label{eq:iso-non-linear}
{\bf H} =w_{iso}'(|B| ) \frac{\bf B}{|B|}.
\end{eqnarray}
The derivative $w_{iso}'$ is exactly the result of measurements of BH-curves. 
\end{itemize}

\section{Lamination energy func\-tion}
\label{sec:lam_energy_function}

In principal, modelling a laminated core would require to resolve the
individual layers of magnetic material and non-magnetic insulation in
the FEA, which is due to the high number of sheets
impractical. Therefore the method for homogenisation is used which
models effective~$\bf B$ and~$\bf H$ fields instead. To derive the
model the following is assumed:
\begin{itemize}
\item The coordinate system is such that the sheets are in $xy$-plane
  and stacked in $z$-direction
\item Material 1 has the properties of vacuum (eq.~\ref{eq:energy_vacuum})
\item The magnetic properties for material 2 is described by a general
  energy density $w_2({\bf B})$.
\item The thickness of the sheets is so small that changes in the
  field from one sheet to its neighbouring sheet can be neglected, and
  in the same way from one insulation layer to the next (the fields in
  the sheet and the insulation, however, can be different)
\end{itemize}
The volume ratios material 1 and 2 are called $f_1$ and $f_2$
respectively. 

From basic Maxwell-equations it is known that at interfaces the normal
components of the flux density is continuous (Quantities with indexes
1 or 2 refer the respective materials, quantities without this
additional index are homogenised values):
\[
B_z:=B_{1z}=B_{2z}
\]
Homogenised flux density components are introduced for the
non-continuous components by using a mixing rule, which is the natural
choice from Maxwell-equations
\begin{eqnarray*}
B_x&:=&f_1 B_{1x}+f_2 B_{2x}\\
B_y&:=&f_1 B_{1y}+f_2 B_{2y}.
\end{eqnarray*}

Now, the homogenised energy density is a mixing of the
energy density of the sublayers. Using $B_{1x}, B_{1y}$ as the yet
unknown flux densities in the insulation layer the energy density can
be written as
\begin{eqnarray*}
\lefteqn{w_{lam}^\ast(B_x,B_y,B_z,B_{1x},B_{1y})=}\\
&\frac{f_1}{2\mu_0}(B_{1x}^2+B_{1y}^2+B_z^2)\\
&+f_2w_2\left(\frac{1}{f_2}(B_x-f_1B_{1x}),\frac{1}{f_2}(B_y-f_1B_{1y}),B_z\right)
\end{eqnarray*}
which is the weighted sum of the energy densities in the two types of
material. The flux density vector components are the values inside the
respective material.
Minimising the energy with respect to $B_{1x},B_{1y}$ gives
\begin{eqnarray*}
w_{lam}({\bf B})
=\min_{B_{1x},B_{1y}}w^\ast_{lam}(B_x,B_y,B_z,B_{1x},B_{1y})\\
\end{eqnarray*}
determines the unknowns $B_{1x}, B_{1y}$ and fixes the energy
density. Evaluating the partial derivatives and equation them to zero,
results in the equations
\begin{eqnarray*}
\frac{B_{1x}}{\mu_0}=&F_x(\frac{1}{f_2}(B_x-f_1B_{1x}),\frac{1}{f_2}(B_y-f_1B_{1y}),B_z)\\
\frac{B_{1y}}{\mu_0}
=&F_y(\frac{1}{f_2}(B_x-f_1B_{1x}),\frac{1}{f_2}(B_y-f_1B_{1y}),B_z),
\end{eqnarray*}
which constitute a system of nonlinear equations for $B_{1x},B_{1y}$.
Deriving with respect to $B_x, B_y,
B_z$ 
defines the magnetic field ${\bf H}$.
\begin{eqnarray*}
  {\bf H}{}={}
\left(\begin{array}{c}0\\0\\\frac{f_1B_{z}}{\mu_0}\end{array}\right)
+F\left(\begin{array}{c}
\frac{1}{f_2}(B_x-f_1B_{1x})\\
\frac{1}{f_2}(B_y-f_1B_{1y})\\
B_z\\
\end{array}\right)
\end{eqnarray*}

Putting the equations together the following set of equation describes
a stack of the laminated electric steel:
\begin{eqnarray*}
\label{eq:lamination}
\left (
\begin{array}{c}H_x\\H_y\\\frac{1}{f_2} \left (H_z-\frac{f_1 B_z}{\mu_0} \right) \end{array}  \right )
=  F \left ( \begin{array}{c}
\frac{1}{f_2} (B_x-f_1 H_x \mu_0) \\
\frac{1}{f_2} (B_y-f_1 H_y \mu_0) \\
B_z \end{array} \right )
\end{eqnarray*}

Given the effective flux density the magnetic field can be computed by
solving this system of non-linear equations. This is now the material
law for a stack of electric steel, if the underlying material law of
the steel type is known (which will be discussed below). For general
functions $F$ these equation cannot be solved analytically.  The
system of equation has to be incorporated into a FEA software in a way
that, every time the system enquires the value of the magnetic field
for a given flux density, the system of equations is solved
iteratively.

One step of approximation is as follows: if it is assumed that material~2 is
strongly ferromagnetic then $B_{1x}$ and $B_{1y}$ can be neglected
with respect to $B_{2x}$ and $B_{2y}$ respectively. The energy density
is then approximated by
\begin{eqnarray*}
w_{lam}^{lin}({\bf B})=\frac{f_1B_z^2}{2\mu_0}
  +f_2w_2(\frac{B_x}{f_2},\frac{B_y}{f_2},B_z).
\end{eqnarray*}
This function can be evaluated without solving an additional nonlinear
equation and the magnetic field derives as:
\begin{eqnarray*}
\left(\begin{array}{c}H_x\\H_y\\H_z\end{array}\right)
=\left(\begin{array}{c}F_{2x}(\frac{B_x}{f_2},\frac{B_y}{f_2},B_z)\\ 
F_{2y}(\frac{B_x}{f_2},\frac{B_y}{f_2},B_z)\\ 
\frac{f_1B_z}{\mu_0}+f_2F_{2z}(\frac{B_x}{f_2},\frac{B_y}{f_2},B_z)
\end{array}\right).
\end{eqnarray*}
The advantage of this approximation for FEA is that it can be applied
without solving additional equations. In practical cases, however, it must be
verified that the accuracy of the approximation is sufficient.

\begin{figure}
  \input{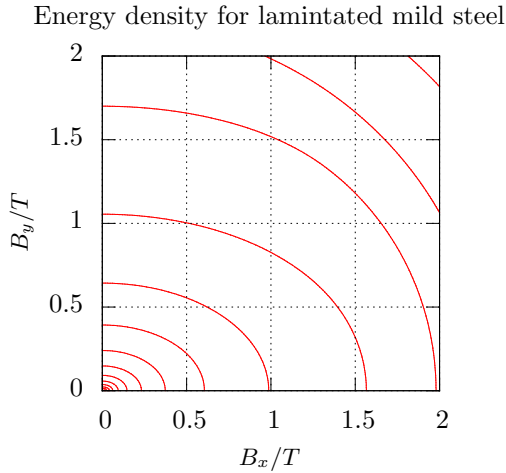}
  \caption{\label{fig:contour_steel_iso_laminate}Contour lines of the
    energy density function for laminated mildsteel.}
\end{figure}

In a first example this theory is applied to mild steel with an
isotropic material behaviour in the manner of eq.~\ref{eq:energy_iso}.
Figure~\ref{fig:contour_steel_iso_laminate} shows the lines of equal
energy that results for a filling factor of $f_2=0.97$.  Compared with
fig.~\ref{fig:contour_steel_iso} - the bulk material - the lines
appear to be squeezed towards the $x$-axis.

\section{Grain-oriented electric steel}
\begin{figure}
\input{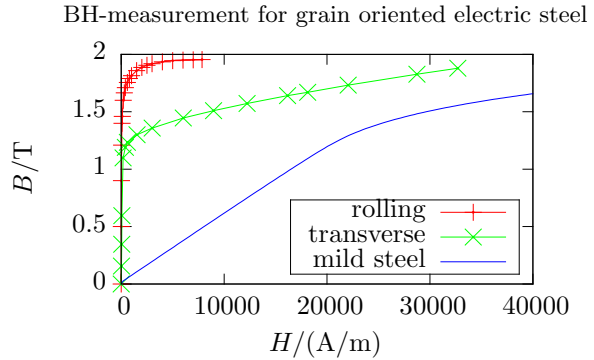}
\caption{\label{fig:bh-measurement}Measurement data for grain-oriented
  electric steel in rolling and in transverse direction compared with
  a measurement of mild steel.}
\end{figure}

Now the method of the energy function shall be applied to model the
magnetic characteristic of grain-oriented electric steel based on
measurement. Typical measurements are along rolling direction and
transverse direction like in fig.~\ref{fig:bh-measurement}. These
measurement are performed using the Epstein-frame by cutting sheets
into stripes along the direction to be measured. In these measurements
the flux density is aligned with the long edge of the stripes and the
magnetic field strength component parallel to that edge is
measured. The stripes are placed in $x$-direction, then the
measurements gives as a function $B\rightarrow H$ that is a part of
the complete three-dimensional function:
\begin{eqnarray*}
  \label{eq:eppstein}
  H_{meas}=H_x(B_{meas},0).
\end{eqnarray*}
In the same way a second measurement along $y$ gives an additional
view.  From these two measurement an interpolation scheme is
established for a two dimensional BH-characteristic. The functions
from measurements will be called $B_0$ and $B_{90}$ for the rolling
direction and the transverse direction, respectively.

First the functions are integrated:
\begin{eqnarray*}
  w_0(B)&=&\int_0^B H_0(B')dB'\\
  w_{90}(B)&=&\int_0^B H_{90}(B')dB'.\\
\end{eqnarray*}
These functions are the basis for an interpolation rule for the energy
density function $w: B_x,B_y\rightarrow w(B_x,B_y)$. The main point is
that the lines of equal energy are constructed for all values of~$w$
which sufficiently defines the function. For a given value~$w$ the
inverse functions of $w_0, w_{90}$ are used to find $B_0, B_{90 } $
such that
\begin{eqnarray*}
  w=w_0(B_0)=w_{90}(B_{90}).
\end{eqnarray*}
The iso-line is defined to be an ellipse:
\begin{eqnarray}
  \left(\frac{B_x}{B_0(w)}\right)^2
  +\left(\frac{B_y}{B_{90}(w)}\right)^2
=1
\end{eqnarray}
Apparently the enclosed area is convex which is essential because a
convex interpolation function must be found. In a practical
implementation the flux density is given and the energy density must
be found. A system of two equations has to be solved in this case. The
two equations are the ellipse equation and the requirement that the half
axes of the ellipse correspond to the same energy. To be solved for
$B_0, B_{90}$:
\begin{eqnarray}
  \left(\frac{B_x}{B_0}\right)^2
  +\left(\frac{B_y}{B_{90}}\right)^2
=1\\
w_0(B_0)=w_{90}(B_{90}).
\end{eqnarray}
From the ellipse equation the equation determining the magnetic field
strength are derived by taking the partial derivatives with respect to
the flux density components:

\begin{equation}
\left( 
\begin{array}{c} H_x\\H_y \end{array}
\right)
=\frac{1}{\frac{B_x^2}{B_0^3H_0}+\frac{B_y^2}{B_{90}^3H_{90}}}
\left( 
\begin{array}{c} B_x/B_0^2\\B_y/B_{90}^2 \end{array}
\right)
\end{equation}
Here only the two-dimensional case is shown, the extension to
three dimensions is obvious and leads to a system of three equations.

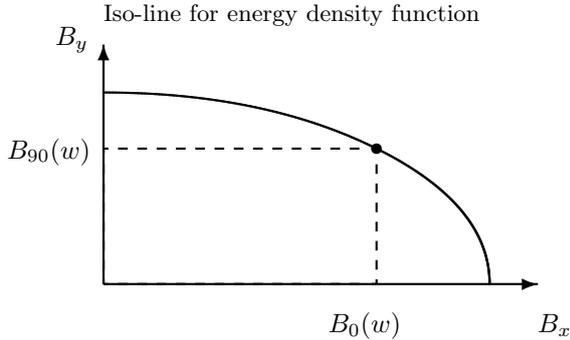
\begin{figure}
\setlength{\unitlength}{0.5in}
%
%
\begin{picture}(6,4)
\thicklines
\put(1,1){\dashbox{0.1}(2.8284,1.4142){}}

\put(1,1){\vector(1,0){4.5}}
\put(1,1){\vector(0,1){2.5}}
\put(5.5,0.5){$B_x$}
\put(0.5,3.5){$B_y$}
\put(1,3.75){\small Iso-line for energy density function}
\put(3.3284,0.5){$B_0(w)$}
\put(0,2.3142){$B_{90}(w)$}

\put(3.8284,2.4142){\circle*{0.1}}

\bezier{215}(5,1)(5,1.8284)(3.8284,2.4142)
\bezier{215}(3.8284,2.4142)(2.6569,3)(1,3)

\thinlines
\end{picture}
\caption{\label{fig:ellipse}The iso-line for energy density $w$ is
  constructed of be an ellipse that meats measured results at the axes.}
\end{figure}
\section{Application example}
\begin{figure}
\input{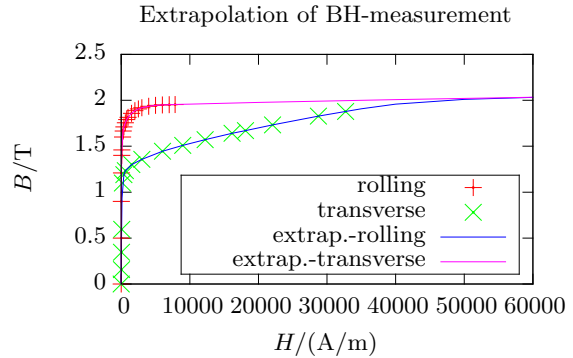}
\caption{\label{fig:bh-measurement2}Measurement data for grain-oriented
  electric steel in rolling and in transverse direction and
  extrapolated data to capture saturation.}
\end{figure}

The interpolation scheme shall be applied to the measured BH-curves
shown in fig.~\ref{fig:bh-measurement}. The measurements are limited to
below 2~T, but when applications in the saturation regime are targeted, 
the measured data must be extrapolated. For the case of the
rolling direction saturation is practically reached and the curve can
be extended using the permeability of one:
\[
B_{high field}=B_{sat}+\frac{H}{\mu_0}.
\]
The case for the transverse direction is more
complicated. Reference~\cite{Jiles}) states that in any direction the
saturation flux density is the same. Therefore a continuation is
constructed that approaches the same saturation equation. This is
depicted in fig.~\ref{fig:bh-measurement2}.

\begin{figure}
\input{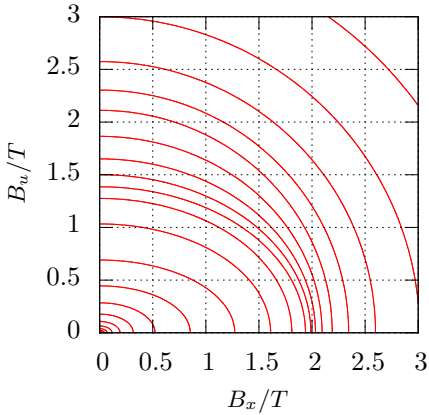}
\caption{\label{fig:bh-go-ansio}Iso-lines of magnetic energy for
  grain-oriented electric steel based on measurements along rolling
  direction($x$) and transverse direction($y$); other points according
  presented interpolation scheme.}
\end{figure}

The application of the interpolation scheme in the sheet plane give
the iso-line plot shown in fig.~\ref{fig:bh-go-ansio}. The ellipses
are clearly anisotropic but tend to become closer to circles at higher
flux densities of $\approx$3~T when saturation is reached and the
material becomes isotropic.

\begin{figure}
\input{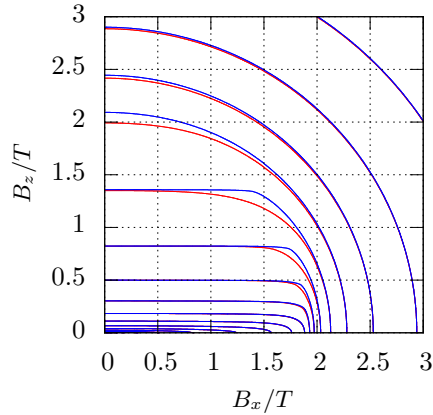}
\caption{\label{fig:bh-go-ansio2}Iso-lines if energy density of grain
  oriented electric steel with lamination effect. Model 1 (red line)
  uses the BH-curve of the transverse direction in perpendicular
  direction; model 2 (blue line) uses the measurement from rolling
  direction. }
\end{figure}

The next step is to combine the result of the grain-oriented material
with the lamination effects. Here measurements for the
BH-characteristic in direction perpendicular to the sheet are needed
in order to apply the interpolation scheme in three dimensions. There
are no such measurements publicly available. 
The interpolation rule with two extreme cases is demonstrated. One 
uses the BH-curve of the rolling direction also for the out-of-plane
direction. The other case uses the measurement in transverse
direction as BH-curve in the perpendicular direction. The two results
for the energy density are both displayed in
fig.~\ref{fig:bh-go-ansio2}. These iso-lines show are cut in the plane
of rolling direction and stacking direction of the three-dimensional
picture. It can be seen that the difference between both cases is
small.  For points with dominating {\em rolling} component the two
models naturally give same results. With dominating {\em stacking}
component the filling factor is determining the result, hence both
models give the similar results.

\section{Conclusion}
This article proposes the use of the energy density function as the
appropriate quantity to describe magnetic constitutive
relations. Energy consumption is guaranteed. The dependence of the
field strength on the flux density is found by derivation. The
differential reluctivity is automatically symmetric.

For a thermodynamically stable material the energy density is a convex
function of the flux density. This leads to a positive definite 
differential reluctivity  and the finite-element-method is  stable.

The method is in particular suitable for an\-iso\-tropic materials. The
BH-law for anisotropy from lamination is demonstrated.
For grain oriented steel an anistropic material law is derived from
two measurements (along rolling direction and transverse
directions). The interpolation for arbitrary directions is done using
the energy density and by building isosurfaces for this function by
using measured date as starting points. This BH-law is used as input
to the previously found lamination law.

If measurements are available in intermediate directions the data can
included in the isosurface construction.

It must be repeated that be presented method is only applicable for
anhysteretic material because only then the energy density is a function
of one variable (the flux density) alone. Furthermore the measurements
must be performed such that the material is always in a thermodynamic
stable state, on then the convexity condition is valid.\\[1cm]

\bibliography{mendeley}

\begin{thebibliography}{10}

\bibitem{BarnerFlohr}
Martin Barner and Friedrich Flohr.
\newblock {\em Analysis II}.
\newblock De Gruyter, 2 edition, 1989.

\bibitem{DiNapoli1983}
Augusto {Di Napoli} and Roberto Paggi.
\newblock {A model of anisotropic grain-oriented steel}.
\newblock {\em IEEE Transactions on Magnetics}, 19(4):1557--1561, July 1983.

\bibitem{Enokizono1997}
Masato Enokizono and Naoya Soda.
\newblock {Finite Element Analysis of Transformer Model Core with Measured
  Reluctivity Tensor}.
\newblock {\em IEEE Transactions on Magnetics}, 33(5):4110--4112, 1997.

\bibitem{Liu1994}
{J. Liu}, A.~Basak, A.~J. Moses, and G.H. Shirkoohi.
\newblock {A method of anisotropic steel modelling using finite element method
  with confirmation by experimental results}.
\newblock {\em IEEE Transactions on Magnetics}, 30(5):3391--3394, 1994.

\bibitem{Janicke1997}
Lutz J\"{a}nicke, Arnulf Kost, Rolf Merte, T.~Nakata, N.~Takahashi,
  K.~Fujiwara, and K.~Muramatsu.
\newblock {Numerical modeling for anisotropic magnetic media including
  saturation effects}.
\newblock {\em IEEE Transactions on Magnetics}, 33(2):1788--1791, March 1997.

\bibitem{Jiles}
David Jiles.
\newblock {\em Introduction to Magnetism and Magnetic Material}.
\newblock Taylor and Francis, 2 edition, 1998.

\bibitem{Kaimori2007}
Hiroyuki Kaimori, Akihisa Kameari, and Koji Fujiwara.
\newblock {FEM Computation of Magnetic Field and Iron Loss in Laminated Iron
  Core Using Homogenization Method}.
\newblock {\em IEEE Transactions on Magnetics}, 43(4):1405--1408, April 2007.

\bibitem{LLem}
Evgeni Mikhailovich~Lifshitz Lev Davidovich~Landau.
\newblock {\em Electrodynamics of continuous media}.
\newblock Butterworth-Heinemann, 1995.

\bibitem{Pera1993}
Thierry Pera, Florence Ossart, and Thierry Waeckerle.
\newblock {Field computation in non linear anisotropic sheets using the
  coenergy model}.
\newblock {\em IEEE Transactions on Magnetics}, 29(6):2425--2427, 1993.

\bibitem{Shirkoohi1994}
G.H. Shirkoohi and J.~Liu.
\newblock {A finite element method for modelling of anisotropic grain-oriented
  steels}.
\newblock {\em IEEE Transactions on Magnetics}, 30(2):1078--1080, March 1994.

\bibitem{Silva1995}
Viviane~Cristine Silva, G\'{e}rard Meunier, and Albert Foggia.
\newblock {3-D Finite-Element Computation of Eddy Currents and Losses in
  Laminated Iron Cores Allowing for Electric and Magnetic Anisotropie}.
\newblock {\em IEEE Transactions on Magnetics}, 31(3):2139--2141, 1995.

\bibitem{Silvester1991}
Peter~P. Silvester and Rajendra~P. Gupta.
\newblock {Effective computational models for anisotropic soft B-H curves}.
\newblock {\em IEEE Transactions on Magnetics}, 27(5):3804--3807, 1991.

\bibitem{VandeSande2004}
Hans VandeSande, Tim Boonen, Ioan Podoleanu, Francois Henrotte, and Kay
  Hameyer.
\newblock {Simulation of a Three-Phase Transformer Using an Improved Anisotropy
  Model}.
\newblock {\em IEEE Transactions on Magnetics}, 40(2):850--855, March 2004.

\end{thebibliography}
\bibliographystyle{hep}

\end{document}